# Majority Based TAS/MRC Scheme in Downlink NOMA Network with Channel Estimation Errors and Feedback Delay

Mahmoud Aldababsa, Oğuz Kucur

*Abstract*— The antenna selection (AS) in non-orthogonal multiple access (NOMA) networks is still a challenging problem since finding optimal AS solution may not be available for all channel realizations and has quite computational complexity when it exists. For this reason, in this paper, we develop a new suboptimal solution, majority based transmit antenna selection (TAS-maj), with significant reduction in computational complexity. The TAS-maj basically selects the transmit antenna with the majority. It is more efficient when compared to previously proposed suboptimal AS algorithms, namely max-max-max ($A^3$) and max-min-max (AIA) because these schemes are merely interested in optimizing the performance of the strongest and weakest users, respectively at the price of worse performance for the remaining users. On the other hand, the TAS-maj scheme yields better performance for more than half of mobile users in the NOMA networks. In this paper, we consider a multiple-input multiple-output communication system, where all the nodes are equipped with multi-antenna. Besides the TAS-maj is employed at the base station, a maximal ratio combining (MRC) is also employed at each mobile user in order to achieve superior performance. The impact of the channel estimation errors (CEEs) and feedback delay (FD) on the performance of the TAS-maj/MRC scheme is studied in the NOMA network over Nakagami-*m* fading channels. The outage behavior of the network is investigated by deriving the exact outage probability (OP) expression in closed-form. In addition, the corresponding upper bound of the OP is obtained in the presence of the CEEs and FD. The OP expression in high signal-to-noise ratio region is also provided to illustrate an error floor value in the presence of the CEEs and FD as well as diversity and array gains in the absence of the CEEs and FD. The analytical results in the presence and absence of the CEEs and FD are verified by the Monte Carlo simulations. The numerical results imply that the system performance is more sensitive to the CEE than FD and shows the superiority of the proposed TAS-maj/MRC scheme over both $A^3$/MRC and AIA/MRC schemes.

## I. INTRODUCTION

Over the last decades, the life has been quite fast taken over by wireless communications (WCs). Just a few years ago, the connectivity of the long-term-evolution (LTE) or what is common as fourth generation (4G) reached to the smart phones and other life applications. The speed of the data transmission is scaled up tens Mbps while the latency of the connection is reduced to 50 milliseconds. A more interesting event, the 5G is coming soon. Its connection speed is expected to reach 10 Gbps, which is about thousand times faster than that of the 4G, making the possibility to download high quality movies/videos in a few seconds for example. The 5G and beyond will allow a high decline in the delay time of the WCs, which is probably 1 millisecond. Besides, a remarkable improvement in mobile coverage is that millions of devices and/or users can be connected simultaneously forming something called Internet of Things (IoT) [1]. And then, it will be expanded to something well-known as Internet of Everything (IoE). Actually, to realize these things, a super multiple access (MA) technique is required. Different from conventional orthogonal multiple access (OMA), which is basically used in the previous generations of WCs (1G-4G), it should have the capability of fulfilling almost all promises of the 5G and beyond. For this reason, non-orthogonal multiple access (NOMA) has been recommended as a promising MA technique for next generations of WCs since it has a variety of significant features that can meet the current emerging demands such as low latency, high spectral efficiency, improved user fairness, and high connection density [2]. In the conventional OMA, the mobile user can only use limited partition of available orthogonal spectrum resources at the transmitter while the users' information signals can be readily separated by taking advantage of the orthogonality at the receiver. However, the key idea of the NOMA is to allocate non-orthogonal resources to serve multiple users, yielding a high spectral efficiency while having controllable interference at the receivers [3]. Looking at NOMA in a little more detail, especially in the power-domain NOMA, multiple users that have differences in power levels can utilize the whole available bandwidth concurrently. At the transmitter side, the information signals of the users are superimposed into one signal. While at the receiver, it is possible to totally isolate the signal of the user with high power level and then cancel it out to leave only the signal of the user with low power level by using successive interference cancellation (SIC) [4].

Recently, multiple-input multiple-output (MIMO) is considered as an extraordinary powerful and useful technique for the WC systems [5]. The gist fundamental of the MIMO is to equip multiple antennas at both transmitter and receiver ends in order to take advantage of the spatial degrees of freedom. It satisfies much higher capacity and better error performance. Unfortunately, undesirable requirements for the signal processing at all transmit and receive nodes are appeared, for instance, considerable amount of power consumption, expensive cost of the radio frequency (RF) chains and excessive computational complexity. However, antenna selection (AS) techniques have been introduced as an impressive solution keeping full diversity order while avoiding extra hardware costs, reducing system complexity and unnecessary power consumption [6]. One common scheme is transmit antenna selection (TAS), which is employed at the transmitter side yielding great reduction in both complexity and power consumption through transmitting all the information over only one RF chain.

In the NOMA, the TAS cannot be carried out directly the same way as in the conventional OMA. It has to be performed in the whole network considering all the mobile users in order to select the transmit antenna with the highest channel gain between the base station (BS) and each mobile user, i.e., the same transmit antenna of the BS must be selected for all the mobile users. Such a transmit antenna may not exist for all channel realizations. Besides this problem, there are severe inter-user interferences in the NOMA networks while the signals are transmitted in an interference free manner in the OMA networks. Therefore, the TAS schemes that can be employed in the conventional OMA networks cannot be extended to the NOMA networks easily. In other words, although the TAS schemes in both the conventional OMA and NOMA networks are similar in principle, due to the interference based on the power domain transmission and reception using the SIC, the mathematical analyses of the NOMA networks using the TAS are more difficult. As a result, because the optimal TAS solution (means to find the best transmit antenna offering the highest channel gain for all the mobile users) cannot be found in all channel realizations and has high computational complexity when it exists, researchers have tried to find different suboptimal TAS solutions with lower computational complexities. Specifically, TAS, and joint TAS and user selection algorithms were proposed in [7] and [8], respectively. Unfortunately, the authors only focused on the design of these algorithms at the BS without any beneficial analytical expression for the system performance. However, in [9], two efficient joint AS algorithms, namely max-max-max ($A^3$) and max-min-max (AIA), are proposed to maximize the sum-rate of the two-user MIMO-NOMA network. In particular, the $A^3$ scheme improves weak user's performance while the AIA enhances strong user's performance. Next, in [10], the same AS algorithms of [9] are again proposed for two different NOMA networks; one is the NOMA with fixed power allocation and the other is the NOMA with cognitive radio-inspired power allocation. Despite the authors in [9] and [10] derived the asymptotic expressions in closed-form for the average rates of both the $A^3$ and AIA in the high signal-to-noise (SNR) region, they did not study the system outage performance. Furthermore, the overall outage probability (OP) expressions are derived in closed-form for these algorithms in [11] and [12]. So far, there are also a few works focused on improving the outage security of the TAS in the NOMA networks in [13]-[15].

Based on the aforementioned literature, the performance of the NOMA with the AS schemes is studied only for two mobile users and only in the ideal case. Moreover, the previously proposed algorithms optimized only the performance of the weak or strong users and this superior performance comes at the price of performance loss for the rest of the mobile users. However, in this paper, we develop a novel majority based TAS (TAS-maj) scheme which aims to optimize the performance of more than half of the mobile users besides a dramatic reduction in computational complexity. Meanwhile, maximal ratio combining (MRC), an optimum diversity technique is applied at the receiver side providing maximum diversity and array gains. Because of these advantages, the TAS-maj can be combined with the MRC scheme in order to have superior performance. To the best of our knowledge, the performance of the hybrid diversity schemes have not been investigated in the NOMA networks in the presence of combined practical impairments of channel estimation errors (CEEs) and feedback delay (FD). With this motivation, we study the performance of the TAS-maj/MRC in the downlink NOMA network in the presence and absence of the CEEs and FD. In this context, the main contributions of this paper can be summarized as follows: (1) The cumulative distribution function (CDF) of the TAS-maj/MRC scheme is obtained. (2) The outage performance of the network over Nakagami-$m$ fading channels in the presence and absence of the CEE and FD is provided. (3) The closed-form expressions of the OP are derived for all mobile users. (4) The upper bound of the OP is obtained in the presence of the CEE and FD. (5) The high SNR analysis is accomplished in order to show the value of error floor (EF) which occurs as a result of the practical impairments. (6) In the ideal case, an asymptotic analysis is carried out such that the achieved diversity and array gains are shown. (7) The theoretical results are validated by the simulations and show the superiority of the proposed TAS-maj/MRC over the previously proposed AS schemes with MRC ($A^3$/MRC and AIA/MRC).

The remainder of this paper is organized as follows: In Section II, the system and channel models are described. In Section II, the TAS-maj/MRC scheme is illustrated. In Section VI, the performance analysis is conducted. The numerical results are presented to verify the analysis in Section V, and the paper is concluded in Section VI.

**Notations:** Throughout this paper, $P_r(.)$ symbolizes probability; $F_X(.)$ denotes the cumulative distribution function (CDF) of a random variable $X$; $E[.]$, $(.)^H$ and $I_N$ represent the expectation operator, the Hermitian transpose and the identity matrix of order $N$, respectively; $\Psi(.,.)$ and $\Gamma(.)$ are lower incomplete Gamma function and Gamma function, respectively; $\|.\|$ denotes Frobenius norm and $J_0(.)$ denotes the zero-order Bessel function of the first kind.

## II. SYSTEM AND CHANNEL MODELS

We consider a downlink MIMO-NOMA network, shown in Fig. 1, where a BS simultaneously serves three mobile users, i.e., $L = 3$. The BS and each of the mobile users are equipped with $N_t = 2$ transmit and $N_r$ receive antennas, respectively. A homogeneous network topology is considered, in which the mobile users are relatively configured in a cluster or close group. The TAS-maj/MRC scheme is applied to the network such that the diversity schemes TAS-maj and MRC are employed at the BS and mobile users, respectively. In the network, the fading channel coefficients between the $i$th antenna of the BS and $j$th receive antenna of the $l$th user ($U_l$) are denoted by $h_l^{(i,j)}$, where $i = 1,2$, $j = 1,\ldots,N_r$ and $l = 1,2,3$. The magnitudes of the fading gains are identically and independently distributed (i.i.d.) Nakagami random variables with fading parameter $m$ and squared mean $\Omega = E[|h_l^{(i,j)}|^2]$. Then, the square of Nakagami-$m$ random variable $|h_l^{(i,j)}|^2$ is Gamma distributed. In terms of the lower incomplete Gamma function [6, eq.(8.350.1)], and Gamma function [6, eq.(8.310.1)], the CDF of the Gamma random variable $|h_l^{(i,j)}|^2$ can be given as $F_{|h_l^{(i,j)}|^2}(x) = \frac{\Psi\left(m,\frac{mx}{\Omega}\right)}{\Gamma(m)}$. We assume that the BS obtains the knowledge of necessary channel state information (CSI) and order of channel quality for the links between the BS and mobile users. During the training period, by assuming the reciprocity of the downlink and uplink channels, each user sends pilot symbols to the BS, then the BS estimates the channel gains and determines index of the majority transmit antenna of the BS, and order of the channel quality of the links. The BS also sends the channel gains to the users through feedback channel. However, in the real communication systems the perfect CSI assumption is not possible due to many practical impairments such as CEEs and inevitable FD. In order to model the system in practical situations, let's define $\boldsymbol{h}_l^{maj}$ as an ($N_r \times 1$) fading channel coefficient vector between the selected majority transmit antenna $i_{maj}$ at the BS and $U_l$ with $\Omega = E[\|\boldsymbol{h}_l^{maj}\|^2]$. Then, by including the CEE, $\boldsymbol{h}_l^{maj}$ can be expressed as $\boldsymbol{h}_l^{maj} = \widehat{\boldsymbol{h}}_l^{maj} + \boldsymbol{e}_{l,cee}$, where $\widehat{\boldsymbol{h}}_l^{maj}$ is the ($N_r \times 1$) estimated channel coefficient vector of $\boldsymbol{h}_l^{maj}$. $\boldsymbol{e}_{l,cee}$ is the ($N_r \times 1$) CEE vector which is statistically independent of $\widehat{\boldsymbol{h}}_l^{maj}$ and modelled as a complex Gaussian random variable [17] with zero mean and $E[\boldsymbol{e}_{l,cee}\boldsymbol{e}_{l,cee}^H] = \boldsymbol{I}_{N_r}\sigma_{e_{l,cee}}^2$. Here, $\sigma_{e_{l,cee}}^2 = \Omega - \widehat{\Omega}$ is the variance of each element in $\boldsymbol{e}_{l,cee}$ and $\widehat{\Omega} = E[\|\widehat{\boldsymbol{h}}_l^{maj}\|^2]$.

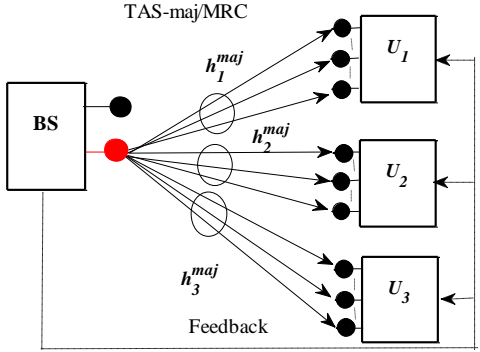

Figure 1: The system model of the downlink NOMA network with the TAS-maj/MRC.

Note that the CEE between the reciproc channels is also included in this model. Meanwhile, when the BS sends the estimated channel gains to the users through the feedback channel, a delay may occur. Consequently, by taking the delay into account, the estimated channel vector can be expressed as $\widehat{\boldsymbol{h}}_l^{maj} = \rho\widehat{\boldsymbol{h}}_l^{(\tau,maj)} + \boldsymbol{e}_{l,fd}$, where $\widehat{\boldsymbol{h}}_l^{(\tau,maj)}$ is the ($N_r \times 1$) FD vector of the estimated channel vector $\widehat{\boldsymbol{h}}_l^{maj}$. Here, $\boldsymbol{e}_{l,fd}$ is the ($N_r \times 1$) FD error vector which is zero mean complex Gaussian random variable with $E[\boldsymbol{e}_{l,fd}\boldsymbol{e}_{l,fd}^H] = \boldsymbol{I}_{N_r}\sigma_{e_{l,fd}}^2$, where $\sigma_{e_{l,fd}}^2$ is the variance of each element in $\boldsymbol{e}_{l,fd}$. Without loss of generality, we can assume that the variances of $\widehat{\boldsymbol{h}}_l^{maj}$ and $\widehat{\boldsymbol{h}}_l^{(\tau,maj)}$ are equivalent [18]. Thus, the $\sigma_{e_{l,fd}}^2 = (1 - \rho^2)\widehat{\Omega}$ [18], where $\rho = J_0(2\pi f_d\tau)$, $0 < \rho < 1$, [19] is the time correlation coefficient, which is defined to qualify the feedback information, and $f_d\tau$ is normalized Doppler frequency. Here, $f_d$ and $\tau$ denote the maximum Doppler frequency of the time-varying channel and the time-delay related to the feedback information, respectively. Since the $\boldsymbol{e}_{l,cee}$ and $\boldsymbol{e}_{l,fd}$ are statistically independent random variables, we can define a new equivalent ($N_r \times 1$) vector, namely $\boldsymbol{e}_l = \boldsymbol{e}_{l,cee} + \boldsymbol{e}_{l,fd}$, which is zero mean complex Gaussian random variable with $E[\boldsymbol{e}_l\boldsymbol{e}_l^H] = \boldsymbol{I}_{N_r}\sigma_{e_l}^2$, where $\sigma_{e_l}^2 = \sigma_{e_{l,cee}}^2 + \sigma_{e_{l,fd}}^2$ is the variance of each element in $\boldsymbol{e}_l$. Then, $\boldsymbol{h}_l^{maj} = \rho\widehat{\boldsymbol{h}}_l^{(\tau,maj)} + \boldsymbol{e}_l$.

Now, in the downlink MIMO-NOMA network, the BS, with transmit power $P_t$, superimposes the mobile users' signals into one signal $x_t = \sum_{i=1}^{L}\sqrt{P_t a_i}x_i$, and then transmits it to all the mobile users. Here $x_i$ is information of $U_i$ with unit energy, and $a_i$ is the power allocation coefficient of the $U_i$ with $\sum_{i=1}^{L}a_i = 1$. The received signal at $U_l$ is expressed as

$$\begin{aligned}\boldsymbol{y}_l &= \boldsymbol{h}_l^{maj}x_t + \boldsymbol{n}_l \\ &= \left(\rho\widehat{\boldsymbol{h}}_l^{(\tau,maj)} + \boldsymbol{e}_l\right)\sum_{i=1}^{L}\sqrt{P_t a_i}x_i + \boldsymbol{n}_l \\ &= \underbrace{\rho\widehat{\boldsymbol{h}}_l^{(\tau,maj)}\sqrt{P_t a_l}x_l}_{\text{Signal term}} + \underbrace{\rho\widehat{\boldsymbol{h}}_l^{(\tau,maj)}\sum_{i\neq l}^{L}\sqrt{P_t a_i}x_i}_{\text{Interference term}} \\ &+ \underbrace{\boldsymbol{e}_l\sum_{i=1}^{L}\sqrt{P_t a_i}x_i}_{\text{Error term}} + \underbrace{\boldsymbol{n}_l}_{\text{Noise term}}, l = 1,2,3. \end{aligned} \quad (1)$$

In (1), the first term is the desired signal of $U_l$, the second term is interference of other users, the third term denotes the error term resulting from CEEs and FD, and the last term indicates the noise such that $\boldsymbol{n}_l$ is the ($N_r \times 1$) zero mean complex additive Gaussian noise with $E[\boldsymbol{n}_l\boldsymbol{n}_l^H] = \boldsymbol{I}_{N_r}\sigma_l^2$ at the $U_l$ and $\sigma_l^2$ is the variance of each element in $\boldsymbol{n}_l$. The received signals are combined by each mobile user with the rule of the MRC. With $\boldsymbol{w}_l = \frac{(\boldsymbol{h}_l^{maj})^H}{\|\boldsymbol{h}_l^{maj}\|} = \frac{\left(\rho\widehat{\boldsymbol{h}}_l^{(\tau,maj)} + \boldsymbol{e}_l\right)^H}{\|\rho\widehat{\boldsymbol{h}}_l^{(\tau,maj)} + \boldsymbol{e}_l\|}$ and $\|\boldsymbol{w}_l\| = 1$, then the combined signal at $U_l$ can be written as

$$\begin{aligned}r_l &= \boldsymbol{w}_l\boldsymbol{y}_l \\ &= \underbrace{\boldsymbol{w}_l\rho\widehat{\boldsymbol{h}}_l^{(\tau,maj)}\sqrt{P_t a_l}x_l}_{\text{Signal term}} + \underbrace{\boldsymbol{w}_l\rho\widehat{\boldsymbol{h}}_l^{(\tau,maj)}\sum_{i\neq l}^{L}\sqrt{P_t a_i}x_i}_{\text{Inter ferenceterm}} \\ &+ \underbrace{\boldsymbol{w}_l\boldsymbol{e}_l\sum_{i=1}^{L}\sqrt{P_t a_i}x_i}_{\text{Error term}} + \underbrace{\boldsymbol{w}_l\boldsymbol{n}_l}_{\text{Noise term}}. \end{aligned} \quad (2)$$

### III. TAS-MAJ/MRC SCHEME

In this section, the majority algorithm in the hybrid diversity TAS/MRC is demonstrated. Moreover, the CDF of the TAS-maj/MRC scheme is derived. Finally, a computational complexity comparison between proposed TAS-maj/MRC and both optimal and suboptimal A$^3$/MRC and AIA/MRC schemes is performed.

#### A. Majority Algorithm in the TAS-maj/MRC Scheme

The Majority rule is generally defined as a decision rule that selects alternatives with a majority, that is, more than half of the votes. It is most often used in influential decision-making systems, e.g., elections. In the election system, many candidates are available for selection by voters, for instance in presidential elections. Subsequently, the winner of the polls should have more than half of the votes. In a similar way, in our system, the first and second transmit antennas at the BS can be considered as candidates, whereas the three mobile users as voters. Thus, the selected transmit antenna must offer maximum channel gains for more than half of the mobile users. As seen, the majority concept can be easily developed in the AS techniques especially, the TAS. The TAS-maj scheme can be realized by first applying a common TAS separately for the mobile users $U_l$, $l = 1,2,3$. Hence, the index of the best transmit antenna $i_l^* \in \{1,2\}$ providing the maximum channel gain between the BS and $U_l$ is selected. Secondly, from the set of $i_l^*$, i.e., $\{i_1^*, i_2^*, i_3^*\}$, the index of the transmit antenna $i_{maj}$ with the majority is chosen. However, in this paper, the TAS-maj/MRC scheme is used such that the transmit antenna $i_l^*$ is selected according to the maximum total of squared channel gains at each mobile user. Consequently, the $i_{maj}$ chosen through this selection criterion is stated as
$$i_{maj} = Maj(i_l^*), l = 1,2,3,$$
$$i_l^* = \arg \max_{i=1,2} \{\varphi_l^i = \parallel \hat{h}_l^i \parallel^2 = \sum_{j=1}^{N_r} |\hat{h}_l^{(i,j)}|^2\}, (3)$$

where $Maj(.)$ is the majority function which determines the $i_{maj}$ through Algorithm I. To be more familiar with the majority Algorithm I, many important guides and definitions should be highlighted before:

• $s$th decision set, where $s = 0,1$: a set that classifies the mobile users according to the majority; first group, $(L-s) = (3-s)$ users selecting majority transmit antenna, i.e., the users have selected the same transmit antenna. Second one, $s$ users selecting non-majority transmit antenna, i.e., the $s$ users have selected different transmit antenna from $(3-s)$ users. Accordingly, the $s$th decision set can be written as $(3-s, s)$. Specifically, in the 0th decision set $(3,0)$, 3 users have selected the same transmit antenna. On the other hand, in the 1st decision set $(2,1)$, 2 users have selected the same transmit antenna and the other user has differed.

• For each $s$th decision set, $n$ subsets of different $i_l^*$ combinations of mobile users are labeled as $(s,n)$, where $n = 1, \ldots, \binom{L}{s} = \binom{3}{s}$. Particularly, in the 0th decision set, only one subset is labeled as $(s,n) = (0,1)$ that corresponds to merely one $i_l^*$ combination of mobile users ($i_1^* = i_2^* = i_3^*$). On the other hand, in the 1st decision set, 3 subsets are labeled as $(s,n) = (1,1),(1,2)$ and $(1,3)$ and may correspond to 3 different $i_l^*$ combinations of mobile users, i.e., $i_1^* \neq (i_2^* = i_3^*), i_2^* \neq (i_1^* = i_3^*), i_3^* \neq (i_1^* = i_2^*)$, respectively. For the sake of clarity, the $s$th decision set and corresponding $n$ subsets labeled as $(s,n)$ for different values of $(s,n)$ are given in Table I.

TABLE I
$s$TH DECISION SET AND CORRESPONDING $n$ SUBSETS LABELED AS $(s,n)$

| $s$th | $(s,n)$ | $i_l^*$ combination of mobile users and its meaning |
|---|---|---|
| 0th $(3,0)$ | $(0,1)$ | $i_1^* = i_2^* = i_3^*$: $U_1, U_2$ and $U_3$ selected the same transmit antenna |
| 1st $(2,1)$ | $(1,1)$ | $i_1^* \neq (i_2^* = i_3^*)$: $U_2$ and $U_3$ selected the same transmit antenna and $U_1$ selected different one |
| | $(1,2)$ | $i_2^* \neq (i_1^* = i_3^*)$: $U_1$ and $U_3$ selected the same transmit antenna and $U_2$ selected different one |
| | $(1,3)$ | $i_3^* \neq (i_1^* = i_2^*)$: $U_1$ and $U_2$ selected the same transmit antenna and $U_3$ selected different one |

• $x^{(s,n)}$: a set associated with the subset $(s,n)$, where it contains the channel gains between the first transmit antenna and $(3-s)$ users selecting majority transmit antenna. Then, the maximum of the $x^{(s,n)}$ is denoted by $X^{(s,n)}$ and can be stated as
$$X^{(s,n)} = \max\{x^{(s,n)} = \{\varphi_l^{i=1}, l \neq k\}\}, \quad (4)$$
where $k$ means the index of the users selecting non-majority transmit antenna.

• $y^{(s,n)}$: a set associated with the subset $(s,n)$, where it contains the channel gains between the second transmit antenna and $(3-s)$ users selecting majority transmit antenna. Then, the maximum of the $y^{(s,n)}$ is denoted by $Y^{(s,n)}$ and can be expressed as
$$Y^{(s,n)} = \max\{y^{(s,n)} = \{\varphi_l^{i=2}, l \neq k\}\}, \quad (5)$$

• $z^{(s,n)}, s \neq 0$: a set associated with the subset $(s,n)$, where it contains the channel gains between the transmit antennas and $s$ users selecting non-majority transmit antenna. Then, the minimum of the $z^{(s,n)}$ is denoted by $Z^{(s,n)}$ and can be written as
$$Z^{(s,n)} = \min\{z^{(s,n)} = \{\varphi_k^{i=1}, \varphi_k^{i=2}\}\}. \quad (6)$$
And for the sake of clarity, the $x^{(s,n)}, y^{(s,n)}, z^{(s,n)}, X^{(s,n)}, Y^{(s,n)}$ and $Z^{(s,n)}$ are given in Table II for different values of $(s,n)$.

TABLE II
$x^{(s,n)}, y^{(s,n)}, z^{(s,n)}, X^{(s,n)}, Y^{(s,n)}$ AND $Z^{(s,n)}$

| $x^{(s,n)}, y^{(s,n)}$ and $z^{(s,n)}$ | $X^{(s,n)}, Y^{(s,n)}$ and $Z^{(s,n)}$ |
|---|---|
| $x^{(0,1)} = \{\varphi_1^{i=1}, \varphi_2^{i=1}, \varphi_3^{i=1}\}$ $y^{(0,1)} = \{\varphi_1^{i=2}, \varphi_2^{i=2}, \varphi_3^{i=2}\}$ | $X^{(0,1)} = \max\{x^{(0,1)}\}$ $Y^{(0,1)} = \max\{y^{(0,1)}\}$ |
| $x^{(1,1)} = \{\varphi_2^{i=1}, \varphi_3^{i=1}\}$ $y^{(1,1)} = \{\varphi_2^{i=2}, \varphi_3^{i=2}\}$ $z^{(1,1)} = \{\varphi_1^{i=1}, \varphi_1^{i=2}\}$ | $X^{(1,1)} = \max\{x^{(1,1)}\}$ $Y^{(1,1)} = \max\{y^{(1,1)}\}$ $Z^{(1,1)} = \min\{z^{(1,1)}\}$ |
| $x^{(1,2)} = \{\varphi_1^{i=1}, \varphi_3^{i=1}\}$ $y^{(1,2)} = \{\varphi_1^{i=2}, \varphi_3^{i=2}\}$ $z^{(1,2)} = \{\varphi_2^{i=1}, \varphi_2^{i=2}\}$ | $X^{(1,2)} = \max\{x^{(1,2)}\}$ $Y^{(1,2)} = \max\{y^{(1,2)}\}$ $Z^{(1,2)} = \min\{z^{(1,2)}\}$ |
| $x^{(1,3)} = \{\varphi_1^{i=1}, \varphi_2^{i=1}\}$ $y^{(1,3)} = \{\varphi_1^{i=2}, \varphi_2^{i=2}\}$ $z^{(1,3)} = \{\varphi_3^{i=1}, \varphi_3^{i=2}\}$ | $X^{(1,3)} = \max\{x^{(1,3)}\}$ $Y^{(1,3)} = \max\{y^{(1,3)}\}$ $Z^{(1,3)} = \min\{z^{(1,3)}\}$ |

• $i_{maj}^{(s,n)}$: the selected majority transmit antenna at $s$th decision set and corresponding subset labeled as $(s,n)$.

---

**Algorithm I: Majority algorithm**

Finding the $i_{maj}$ depends on the values of the elements in the subset labeled as $(s,n)$ and arranged into two cases according to the $s$th decision set:

**Case 1** 0th decision set and $(s,n) = (0,1)$: By using Table II, the index of the selected transmit antenna is expressed as
$$i_{maj}^{(0,1)} = \arg \max_{i=1,2} \{X^{(0,1)}, Y^{(0,1)}\}. \quad (7)$$
Here, there is no user selecting non-majority transmit antenna.

**Case 2** 1st decision set and $(s,n) = (1, n \in \{1,2,3\})$: By using Table II, the index of the transmit antenna selected with the majority is found as
if the $Z^{(1,n)}$ is at the first transmit antenna, then
$$i_{maj}^{(1,n)} = \arg \max_{i=1,2} \{X^{(1,n)}, \max(Y^{(1,n)}, Z^{(1,n)})\}. \quad (8)$$
if the $Z^{(1,n)}$ is at the second transmit antenna, then
$$i_{maj}^{(1,n)} = \arg \max_{i=1,2} \{\max(X^{(1,n)}, Z^{(1,n)}), Y^{(1,n)}\}. \quad (9)$$

---

*B. CDF of TAS-maj/MRC Scheme*

Using (3) together with the assumption of i.i.d. random variables, the CDF of the $\varphi = \varphi_l^i = \parallel \hat{h}_l^i \parallel^2$ can be written as
$$F_\varphi(x) = \frac{\Psi\left(mN_r, \frac{mx}{\bar{\Omega}}\right)}{\Gamma(mN_r)}. \quad (10)$$
According to the NOMA concept and without loss of generality, the channel gains are ordered as $\parallel \hat{h}_1^{maj} \parallel^2 \leq \parallel \hat{h}_2^{maj} \parallel^2 \leq \parallel \hat{h}_3^{maj} \parallel^2$ such that the power coefficients are allocated opposite to the order of the channel gains, i.e., $a_1 \geq a_2 \geq a_3$ with $\sum_{i=1}^3 a_i = 1$. Then, simply $\varphi_l^{maj} = \varphi_l^{i_{maj}} = \parallel \hat{h}_l^{maj} \parallel^2, l = 1,2,3$.

**Proposition 1:** The CDF $F_{\varphi_l^{maj}}(x)$ can be written in general form as
$$F_{\varphi_l^{maj}}(x) = \sum_{q=1}^{N_t L} \zeta_{(l,q)} \big(F_\varphi(x)\big)^q, l = 1,2,3, \quad (11)$$
where $\zeta_{(1,1)} = \zeta_{(1,2)} = \frac{3}{2}, \zeta_{(1,3)} = -\frac{9}{2}, \zeta_{(1,4)} = \frac{15}{4}, \zeta_{(1,5)} = -\frac{3}{2}, \zeta_{(1,6)} = \frac{1}{4}, \zeta_{(2,3)} = 3, \zeta_{(2,4)} = -\frac{3}{4}, \zeta_{(2,5)} = -\frac{9}{4}, \zeta_{(2,6)} = 1, \zeta_{(3,5)} = \frac{3}{2}, \zeta_{(3,6)} = -\frac{1}{2}$ and otherwise equal zero.
*See the proof in Appendix*

### C. Computational Complexity of TAS-maj/MRC Scheme

The complexity for the optimal TAS/MRC algorithm is given as $O(N_t N_r^3)$ by hypothetically assuming that optimal solution always exists. Also for both A $^3$/MRC and AIA/MRC algorithms, it can be stated as $O(3N_t N_r + 4N_t)$. Finally, for the proposed TAS-maj/MRC, since TAS/MRC is applied first for the three users separately, and then the index of the majority transmit antenna is found, the complexity of the TAS-maj/MRC can be written as $O(3N_t N_r + 12N_t)$. Now, for the sake of simplicity, assume that $N_t = N_r = N$. Accordingly, the complexities of the optimal TAS, A $^3$, AIA and TAS-maj with MRC are given as $O(N^4)$, $O(3N^2 + 4N)$, $O(3N^2 + 4N)$ and $O(3N^2 + 12N)$, respectively. Note that the suboptimal selection solutions can remarkably reduce the computational complexity compared to the optimal one. Although, the complexities of A $^3$/MRC and AIA/MRC are on the same order as but lower than that of the proposed TAS-maj/MRC, the TAS-maj/MRC provides better performance than A $^3$/MRC and AIA/MRC schemes for two users out of three users.

## IV. PERFORMANCE ANALYSIS

### A. SINR

The error term in (2) that results from CEEs and FD is treated as noise as in [20]. Thus, *the expression of the instantaneous SINR of $U_l$ to detect $U_j$ can be derived as*
$$SINR_{j\to l} = \frac{a_j P_t \rho^2 \varphi_l^{maj}}{P_t \rho^2 \varphi_l^{maj} \sum_{i=j+1}^L a_i + \sigma_e^2 P_t \sum_{i=1}^L a_i + \sigma^2}$$
$$= \frac{a_j \gamma \varphi_l^{maj}}{\gamma \varphi_l^{maj} \sum_{i=j+1}^L a_i + \beta}, j \ne L, j < l, \quad (12)$$
where $\varphi_l^{maj} = \| \hat{\mathbf{h}}_l^{maj} \|^2 = \| \hat{\mathbf{h}}_l^{(\tau,maj)} \|^2$, $\gamma = \frac{P_t}{\sigma^2}$ is the SNR. For mathematical tractability, $\sigma_{e_l}^2 = \sigma_e^2$ and $\sigma_l^2 = \sigma^2$ are assumed. Also, $\beta = \sigma_e^2 \frac{\gamma}{\rho^2} + \frac{1}{\rho^2}$. Next, the SINR for $U_l$ to decode its own signal can be given by $SINR_l = \frac{a_l \gamma \varphi_l^{maj}}{\gamma \varphi_l^{maj} \sum_{i=l+1}^L a_i + \beta}, l \ne L$. And the SNR for the $U_L$ is expressed as $SNR_L = \frac{a_L \gamma \varphi_l^{maj}}{\beta}$.

### B. OP

Using (12), *the OP of the $U_l$ can be given as* $OP_l = P_r\big(SINR_{j\to l} < \gamma_{th_j}\big) = P_r\left(\frac{a_j \gamma \varphi_l^{maj}}{\gamma \varphi_l^{maj} \sum_{i=j+1}^L a_i + \beta} < \gamma_{th_j}\right)$
$$= P_r\big(\varphi_l^{maj} < \beta \theta_l^*\big)$$
$$= F_{\varphi_l^{maj}}(\beta \theta_l^*), l = 1,2,3, \quad (13)$$
where $\theta_l^* = \max_{1 \le j \le l}\{\frac{\gamma_{th_j}}{\gamma(a_j - \gamma_{th_j} \sum_{i=j+1}^L a_i)}\}$ and $\gamma_{th_j}$ denotes the threshold value of the SNR for the $U_j$.

**Proposition 2:** *By using (10) and (11) with (13) and applying series and binomial expansions, the OP of the $U_l$ in the presence of the CEEs and FD is expressed in closed form as*
$$OP_l = \sum_{q=1}^{N_t L} \zeta_{(l,q)} \left(\frac{\Psi\left(mN_r, \frac{m\beta \theta_l^*}{\hat{\Omega}}\right)}{\Gamma(mN_r)}\right)^q$$
$$= \sum_{q=1}^{N_t L} \zeta_{(l,q)} \left(1 - e^{-\frac{m\beta \theta_l^*}{\hat{\Omega}}} \sum_{k=0}^{mN_r-1} \left(\frac{m\beta \theta_l^*}{\hat{\Omega}}\right)^k \frac{1}{k!}\right)^q$$
$$= \sum_{q=1}^{N_t L} \zeta_{(l,q)} \sum_{p=0}^{q} \sum_{k=0}^{p(mN_r-1)} \binom{q}{p}(-1)^p \vartheta_k(p, mN_r)$$
$$\times (\beta \theta_l^*)^k e^{-\frac{pm\beta \theta_l^*}{\hat{\Omega}}}. \quad (14)$$

In (14), $\vartheta_a(b, g_c)$ denotes multinomial coefficients which can be defined as [6, eq.(0.314)], $\vartheta_a(b, g_c) = \frac{1}{ad_0}\sum_{w=1}^{a}(w(b+1) - a)d_w \vartheta_{a-b}(b, g_c), a \ge 1$, where, $d_w = (g_c/\Omega_c)^w/w!$, $\vartheta_0(b, g_c) = 1$, and $\vartheta_a(b, g_c) = 0$ if $w > g_c - 1$. *In the ideal case, $f_d \tau = 0$ and $\sigma_{e,cee}^2 = 0$, i.e., $\rho = 1$ and $\sigma_e^2 = 0$ which give $\beta = 1$, and then by substituting $\beta$ into (14), the OP of the $U_l$ in the absence of the CEEs and FD can be easily obtained.*

### C. OP analysis in high SNR regime

Using high SNR approximation technique proposed in [21], i.e., $\gamma \to \infty$, then $\theta_l^* \to 0$. In addition, by the property of incomplete Gamma function $\Psi(x, y \to 0) \approx y^x/x$ [6, eq. (45.9.1)], the expression in (14) can be expressed as
$$\frac{\Psi\left(mN_r, \frac{m\theta_l^*}{\hat{\Omega}}\right)}{\Gamma(mN_r)}\big|_{(\beta=1,\theta_l^* \to 0)} = \frac{(m/\Omega)^{mN_r}}{\Gamma(mN_r+1)}(\theta_l^*)^{mN_r}$$
$$\le \frac{\Psi\left(mN_r, \frac{m\beta \theta_l^*}{\hat{\Omega}}\right)}{\Gamma(mN_r)}\big|_{(\beta \ge 1, \theta_l^* \to 0)}$$
$$= \frac{(m/\hat{\Omega})^{mN_r}}{\Gamma(mN_r+1)}(\beta \theta_l^*)^{mN_r}, \quad (15)$$

**Proposition 3:** Using (15), *the upper bound of the OP for the $U_l$ in the presence of the CEEs and FD can be stated as*
$$OP_l^{upper} = \sum_{q=1}^{N_t L} \zeta_{(l,q)} \frac{(m/\hat{\Omega})^{qmN_r}}{\Gamma(mN_r+1)^q}(\beta \theta_l^*)^{qmN_r} \quad (16)$$
If $\beta = 1$ in (16), then *the asymptotic OP of the $U_l$, $OP_l^{asymp}$, in the absence of the CEE and FD is obtained*. The $OP_l^{asymp}$ depends on the $\gamma$. Therefore, the diversity and array gains of the system can be achieved in the absence of the CEE and FD. However, in (16), suppose that $\alpha_l = \max_{1 \le j \le l}\{\frac{\gamma_{th_j}}{a_j - \gamma_{th_j} \sum_{i=j+1}^L a_i}\}$ is defined. Thus, $\theta_l^* = \frac{\alpha_l}{\gamma}$ and in the high SNR region, $\beta \theta_l^* = \sigma_e^2 \frac{\gamma}{\rho^2} \frac{\alpha_l}{\gamma} + \frac{1}{\rho^2} \frac{\alpha_l}{\gamma} \to \sigma_e^2 \frac{\alpha_l}{\rho^2}$. Then, in the presence of the CEEs and FD, *the upper EF for the lth user in the high SNR region can be obtained as*
$$EF_l^{upper} = \sum_{q=1}^{N_t L} \zeta_{(l,q)} \frac{(m/\hat{\Omega})^{qmN_r}}{\Gamma(mN_r+1)^q}\left(\sigma_e^2 \frac{\alpha_l}{\rho^2}\right)^{qmN_r}. \quad (17)$$
As observed in (17), EF bounds do not depend on the $\gamma$. Therefore, the OP reaches a fixed value in the high SNR

region, which implies that the presence of CEEs and FD eliminates the diversity gain of the system.

## V. NUMERICAL RESULTS

This section illustrates the effects of CEE and FD in terms of $\sigma_{e,cee}^2$ and $f_d\tau$ $(\rho)$, the number of receive antennas of the mobile users $(N_r)$ and the channel condition $(m)$ on the performance of the MIMO-NOMA network with TAS-maj/MRC. Assume three users with power factors $a_1 = 0.6$, $a_2 = 0.3$ and $a_3 = 0.1$. Also, the corresponding threshold SINR values of the three users can be considered as $\gamma_{th_1} = 1.4$, $\gamma_{th_2} = 2.2$ and $\gamma_{th_3} = 2.5$, and $\Omega = 1$.

Performance of the OP versus SNR is shown in Fig. 2 and Fig. 3 in the presence and absence of CEE&FD, respectively. The OP performance is plotted for $(m; N_t, N_r) = (1; 2,1)$, which is a special case of Rayleigh fading and TAS-maj scheme. The parameters $(f_d\tau, \sigma_{e,cee}^2)$ are considered as $(0.01, 0.001)$ and $(0,0)$ for the presence and absence of CEE&FD, respectively. As seen in Fig. 2 and Fig. 3, the analytical and simulation results match perfectly. Moreover, the A$^3$ and AIA schemes enhance the strong user ($U_3$) and weak user ($U_1$), respectively and the other users have worse performance. However, the proposed TAS-maj scheme improves the performance of two users for both cases of the presence and absence of CEE and FD, and keeps reasonable good performance for the other. In other words, the TAS-maj scheme provides better performance than the A$^3$ scheme for two users ($U_1$ and $U_2$) and AIA scheme for two users ($U_2$ and $U_3$). Note that the proposed scheme outperforms the other schemes significantly for the $U_2$, e.g., at an OP of $10^{-4}$, the proposed one accomplishes approximately 7.5 dB SNR gain as compared to the other schemes in absence of CEE and FD. Note also that the presence of CEE and FD causes an error floor in the high SNR region.

In order to show the superiority of the proposed diversity scheme TAS-maj/MRC over A$^3$/MRC and AIA/MRC schemes for more than three users, performance of the OP for five mobile users is simulated in Fig. 4 and Fig. 5, respectively. The power factors of the users are assumed as $a_1 = 9/20$, $a_2 = 5/20$, $a_3 = 3/20$, $a_4 = 2/20$ and $a_5 = 1/20$ and the corresponding threshold SINR values of the five users can be considered as $\gamma_{th_1} = 0.8$, $\gamma_{th_2} = 0.8$, $\gamma_{th_3} = 0.9$, $\gamma_{th_4} = 1.5$ and $\gamma_{th_5} = 2$. The OP performance is plotted for $(m; N_t, N_r) = (1; 2,2)$ and the parameters $(f_d\tau, \sigma_{e,cee}^2)$ are considered as $(0.01, 0.01)$ and $(0,0)$ for the presence and absence of CEE&FD, respectively. As seen in Fig. 4, the proposed TAS-maj/MRC scheme provides better performance than the A$^3$/MRC scheme for four mobile users ($U_1$, $U_2$, $U_3$ and $U_4$) and remarkable improvement in the performance for the three mobiles users that located between the weakest and strongest users ($U_2$, $U_3$ and $U_4$). On the other hand, as observed in Fig. 5 the proposed TAS-maj/MRC scheme provides better performance than the AIA/MRC scheme for three mobile users ($U_3$, $U_4$ and $U_5$) in the presence and absence of CEE and FD. However, in the absence of CEE and FD, the proposed scheme provides almost the same performance as the AIA/MRC scheme for $U_2$. Note that Fig. 4 and Fig. 5 reveal that the proposed scheme provides better performance than A$^3$/MRC and AIA/MRC schemes for more than $\frac{L}{2}$ mobile users generally.

Performance of the OP versus SNR is presented in Fig. 6 and Fig. 7 for various antenna configurations and Nakagami parameters in the presence and absence of CEE&FD, respectively such that $(f_d\tau, \sigma_{e,cee}^2)$ are given as $(0.02, 0.01)$ and $(0,0)$, respectively. Note that an excellent agreement between the exact analytical results and simulations is observed. Also, the OP performance improves as the number of antennas increases (superior performance of TAS-maj/MRC to TAS-maj scheme) as well as the OP performance enhances as the channel conditions (values of Nakagami parameter) improve. As seen in Fig. 7, the configuration $(m; N_t, N_r) = (1; 2,2)$ provides approximately 12.5 dB SNR gain at an OP of $10^{-6}$ for the $U_2$ as compared to $(1; 2,1)$ as well as the configuration $(m; N_t, N_r) = (2; 2,2)$ provides approximately 5 dB SNR gain as compared to $(1; 2,2)$. Besides, the asymptotic curves are compatible with the exact ones, and the diversity and array gains are achieved. The diversity gains of the $U_1$ for example, according to the configurations $(m; N_t, N_r) = (1; 2,1), (1; 2,2)$ and $(2; 2,2)$ are 1, 2 and 4, respectively. And for the sake of clarity, the corresponding upper bound of OP$_3$ is shown in Fig. 8. It is noticed that the OP reaches an EF due to CEE and FD when the system SNR increases, which means zero diversity order.

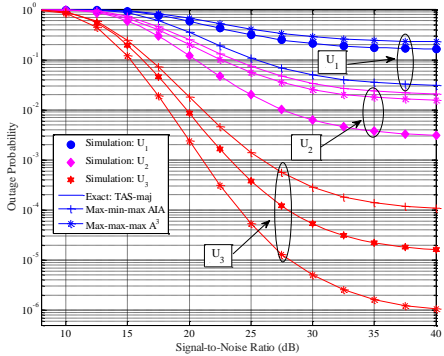

Figure 2: The OP of the TAS-maj-NOMA system versus the SNR in presence of CEE and FD.

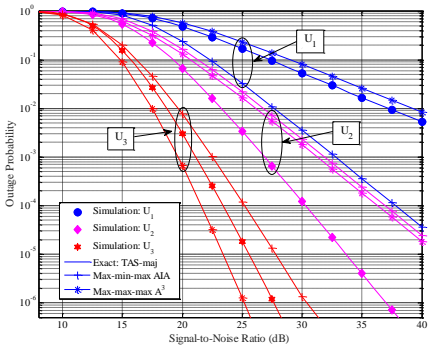

Figure 3: The OP of the TAS-maj-NOMA system versus the SNR in absence of CEE and FD.

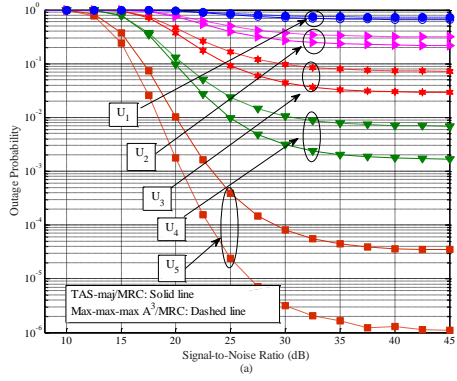

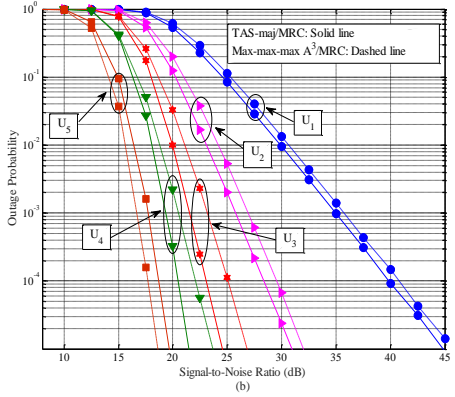

Figure 4: The OP of the TAS-maj/MRC-NOMA system versus the SNR in (a) presence (b) absence of CEE and FD.

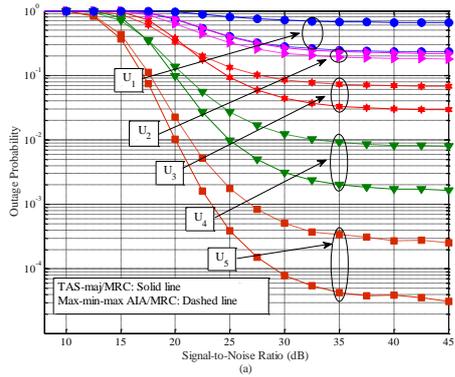

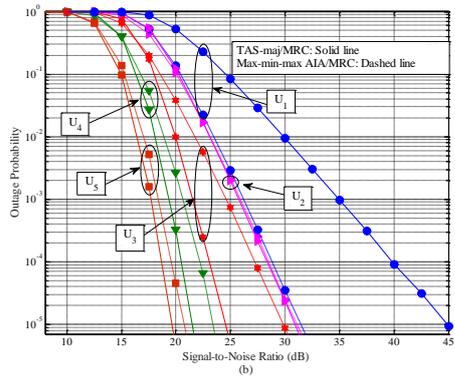

Figure 5: The OP of the TAS-maj/MRC-NOMA system versus the SNR in (a) presence (b) absence of CEE and FD.

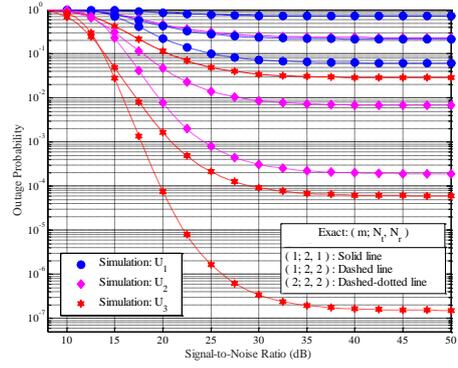

Figure 6: The OP of the TAS-maj/MRC-NOMA system versus the SNR in presence of CEE and FD.

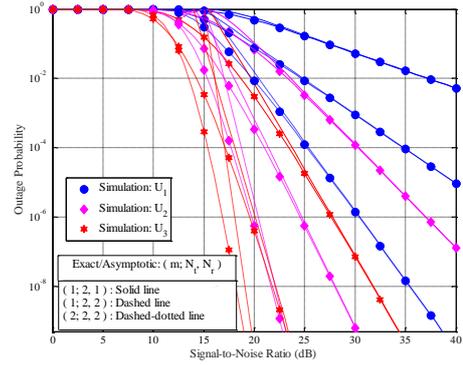

Figure 7: The OP of the TAS-maj/MRC-NOMA system versus the SNR in absence of CEE and FD.

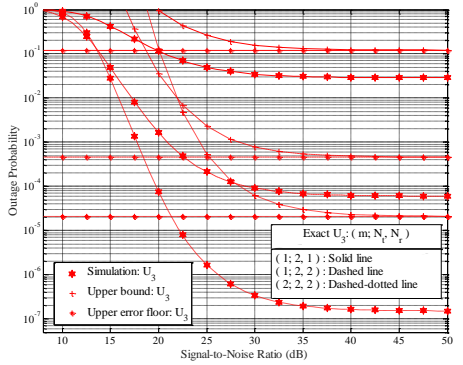

Figure 8: The OP of the TAS-maj/MRC-NOMA system versus the SNR in presence of CEE and FD.

Fig. 9 and Fig. 10 illustrate the performance of the OP versus $\rho$ and $\sigma_{e,cee}^2$, respectively. In Fig. 9, the parameters $(m; N_t, N_r) = (2; 2,3)$, SNR = 15 dB and $\sigma_{e,cee}^2 = 0.02$ are assumed. Whereas in Fig. 10, the parameters $(m; N_t, N_r) = (1; 2,2)$, SNR = 20 dB and $f_d\tau = 0.01$. For fixed CEE in Fig. 9, the performance gets better with shorter FD (higher $\rho$). On the other hand, for fixed FD in Fig. 10, the performance becomes worse as the CEE increases. Furthermore, both Fig. 9 and Fig. 10 show that the proposed TAS-maj/MRC guarantees better performance than A$^3$/MRC and AIA/MRC schemes for two users. However, for any user all schemes converge to the same OP as FD decreases in Fig. 9 and CEE increases in Fig. 10, respectively. Fig. 11 demonstrates the OP performance of the $U_3$ versus the SNR in different cases of $(f_d\tau, \sigma_{e,cee}^2)$ labeled as ideal case $(0,0)$, only FD $(\{0.01,0.02\},0)$, only CEE $(0,0.01)$ and combined CEE&FD $(\{0.01,0.02\},0.01)$. Also, $(m; N_t, N_r) = (1; 2,2)$ is assumed.

Considering ideal case as a reference, i.e., $(f_d\tau, \sigma_{e,cee}^2) = (0,0)$, at an OP of $10^{-4}$ with $f_d\tau = 0.01$, the performance degradations of the proposed scheme in cases of only FD (0.01,0) and combined CEE&FD (0.01,0.01) are approximately 0.5 dB and 5 dB, respectively. On the other hand, with $f_d\tau = 0.02$, the performance degradations of the proposed scheme in cases (0.02,0) and (0.02,0.01) are approximately 3 dB and 14 dB, respectively. Thus, there are 2.5 dB and 9 dB SNR degradations for the cases of only FD and combined CEE&FD, respectively when $f_d\tau$ increases from 0.01 to 0.02. Also, the performance degradation of the proposed scheme between the cases (0,0) and (0,0.01) is approximately 4 dB. These imply that the system performance is more sensitive to CEE than FD. Note that this comment is also valid for the other users, however for the sake of the clarity we only plotted the OP for the third user.

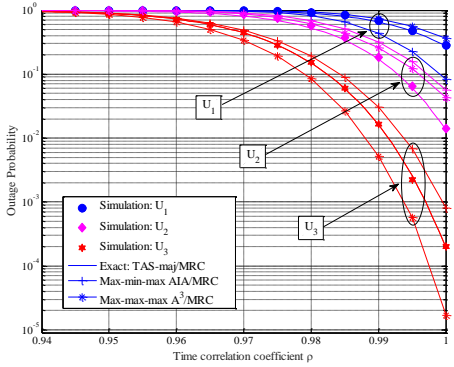

Figure 9: The OP of the TAS-maj/MRC-NOMA system versus the $\rho$.

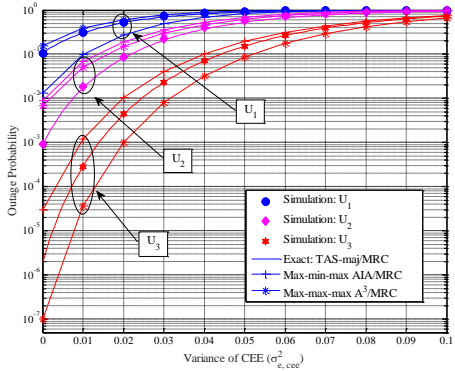

Figure 10: The OP of the TAS-maj/MRC-NOMA system versus the $\sigma_{e,cee}^2$ in presence of CEE and FD.

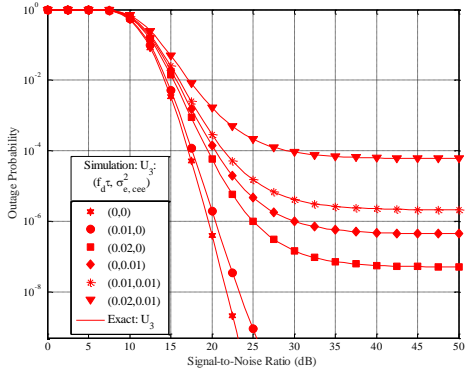

Figure 11: The OP of the TAS-maj/MRC-NOMA system versus the SNR.

## VI. CONCLUSION

In this paper, majority based TAS algorithm has been developed and the outage behavior of the TAS-maj/MRC in the downlink MIMO-NOMA network has been analyzed over Nakagami-$m$ fading channels in the presence and absence of both the CEEs and FD. The closed-form expression of the exact OP is attained. Also, the upper bound of the OP is achieved in the presence of the CEEs and FD. In the high SNR region, the OP reaches the EF due to the presence of the CEEs and FD, and hence the diversity order diminishes while in the absence of the CEEs and FD, better diversity and array gains are achieved. Furthermore, the outage performance of the network becomes remarkable as the channel conditions improve or number of receive antennas increases. Moreover, the system performance is more sensitive to CEE than FD. Finally, in general the proposed TAS-maj/MRC scheme provides better performance than the $A^3$/MRC and AIA/MRC schemes for more than half of the mobile users $\left(>\frac{L}{2}\right)$. Note that the algorithm can be generalized to other cases easily although it has been analyzed for two transmit antennas and three users, aiming simplicity.

## APPENDIX

*Proof of Proposition 1:* Using (10) and with $Q_{l,s} = \frac{(L-s)!}{(L-l)!(l-s-1)!}$, the CDF of the $l$th ordered random variable at the $s$th decision set, $F_{\tilde{\varphi}_{l,s}}(x)$, can be expressed as [23]

$$F_{\tilde{\varphi}_{l,s}}(x) = Q_{l,s}\sum_{t=0}^{L-l}\frac{(-1)^t}{l-s+t}\binom{L-l}{t}\times[F_\varphi(x)]^{l-s+t} \quad (18)$$

Now, in order to find the CDF of the channel gains, $\varphi_l^{maj} = \|\hat{h}_l^{maj}\|^2, l = 1,2,3$, for the mobile users at the $i_{maj}$, we firstly should find the CDFs of $Z^{(s,n)}$ and the $l$th ordered $X^{(s,n)}|_l$ and $Y^{(s,n)}|_l$. Hence, using (4), (5) and (6) and by taking the ordered random variables in the sets $x^{(s,n)}$ and $y^{(s,n)}$, the CDFs' expressions $F_{X^{(s,n)}|_l}(x)$, $F_{Y^{(s,n)}|_l}(x)$ and $F_{Z^{(s,n)}}(x)$ can be stated, respectively as

$$F_{X^{(s,n)}|_l}(x) = F_{\tilde{\varphi}_{l,s}^{i=1}}(x) = F_{\tilde{\varphi}_{l,s}}(x), s = 0,1, \quad (19)$$

$$F_{Y^{(s,n)}|_l}(x) = F_{\tilde{\varphi}_{l,s}^{i=2}}(x) = F_{\tilde{\varphi}_{l,s}}(x), s = 0,1, \quad (20)$$

$$F_{Z^{(s,n)}}(x) = 1 - \left(1 - F_\varphi(x)\right)^{2s}$$
$$= 2F_\varphi(x) - F_\varphi^2(x), s = 1. \quad (21)$$

Next, the channel gains $\varphi_l^{maj}, l = 1,2,3$ and their CDF $F_{\varphi_l^{maj}}(x)$ can be derived as in (22), (23) and (24), respectively, given at the top of next page.

$$\varphi_{3,s}^{maj} = \begin{cases} \max(X^{(0,1)}|_{l=3}, Y^{(0,1)}|_{l=3}) = \max(\tilde{\varphi}_{3,0}^{i=1}, \tilde{\varphi}_{3,0}^{i=2}), & s=0, n=1 \\ \max(X^{(1,n)}|_{l=3}, Y^{(1,n)}|_{l=3}, Z^{(1,n)}) = \max(\tilde{\varphi}_{3,1}^{i=1}, \tilde{\varphi}_{3,1}^{i=2}, Z^{(1,n)}), & s=1, n=1,2,3 \end{cases}$$ (22)

$$\varphi_{2,s}^{maj} = \begin{cases} \max(x^{(0,1)}|_{l=2}, y^{(0,1)}|_{l=2}) = \max(\tilde{\varphi}_{2,0}^{i=1}, \tilde{\varphi}_{2,0}^{i=2}), & s=0, n=1 \\ \max(x^{(1,n)}|_{l=2}, y^{(1,n)}|_{l=2}, z^{(1,n)}) = \max(\tilde{\varphi}_{2,1}^{i=1}, \tilde{\varphi}_{2,1}^{i=2}, z^{(1,n)}), & s=1, n=1,2,3 \end{cases}$$ (23)

$$\varphi_{1,s}^{maj} = \begin{cases} \max(x^{(0,1)}|_{l=1}, y^{(0,1)}|_{l=1}) = \max(\tilde{\varphi}_{1,0}^{i=1}, \tilde{\varphi}_{1,0}^{i=2}), & s=0, n=1 \\ Z^{(1,n)}, & s=1, n=1,2,3 \end{cases}$$ (24)

For $\varphi_3^{maj}$ and $F_{\varphi_3^{maj}}(x)$: The $\varphi_3^{maj}$ can be simply found as the 3rd order channel gain in the sets labeled as $(s,n)$. Thus, the $\varphi_3^{maj}$ in the $s$th decision set, $\varphi_{3,s}^{maj}$, can be stated as in (22), where $\max(\max(X^{(s,n)}|_l, Z^{(s,n)}), Y^{(s,n)}|_l) = \max(X^{(s,n)}|_l, \max(Y^{(s,n)}|_l, Z^{(s,n)})) = \max(X^{(s,n)}|_l, Y^{(s,n)}|_l, Z^{(s,n)})$ is used. Then, by using (18)-(21), the CDFs $F_{\varphi_{3,0}^{maj}}(x)$ and $F_{\varphi_{3,1}^{maj}}(x)$ can be obtained, respectively as

$$F_{\varphi_{3,0}^{maj}}(x) = F_{X^{(0,1)}|_{l=3}}(x) F_{Y^{(0,1)}|_{l=3}}$$
$$= F_{\tilde{\varphi}_{3,0}}(x) F_{\tilde{\varphi}_{3,0}}(x)$$
$$= \left( \frac{3!}{(3-3)!(3-1)!} \sum_{t=0}^{3-3} \frac{(-1)^t}{3+t} \binom{3-3}{t} F_\varphi(x)^{3+t} \right)^2$$
$$= F_\varphi^6(x) \qquad (25)$$

$$F_{\varphi_{3,1}^{maj}}(x) = F_{X^{(1,n)}|_{l=3}}(x) F_{Y^{(1,n)}|_{l=3}}(x) F_{Z^{(1,n)}}(x)$$
$$= F_{\tilde{\varphi}_{3,1}}(x) F_{\tilde{\varphi}_{3,1}}(x) \left( 2F_\varphi(x) - F_\varphi^2(x) \right)$$
$$= \left( F_\varphi(x) \right)^4 \left( 2F_\varphi(x) - F_\varphi^2(x) \right)$$
$$= 2F_\varphi^5(x) - F_\varphi^6(x), \qquad (26)$$

And then, with $P_r(s=0) = \frac{\binom{3}{0}}{\binom{3}{0}+\binom{3}{1}} = \frac{1}{4}$, $P_r(s=1) = \frac{\binom{3}{1}}{\binom{3}{0}+\binom{3}{1}} = \frac{3}{4}$, the CDF $F_{\varphi_3^{maj}}(x)$ can be expressed as

$$F_{\varphi_3^{maj}}(x) = P_r(s=0) F_{\varphi_{3,0}^{maj}}(x) + P_r(s=1) F_{\varphi_{3,1}^{maj}}(x)$$
$$= \frac{3}{2} F_\varphi^5(x) - \frac{1}{2} F_\varphi^6(x). \qquad (27)$$

For $\varphi_2^{maj}$ and $F_{\varphi_2^{maj}}(x)$: After determining $F_{\varphi_3^{maj}}(x)$, there is no necessity to use $X^{(s,n)}|_l$, $Y^{(s,n)}|_l$ and $Z^{(s,n)}$ when finding $\varphi_2^{maj}$. Thus, for the 2nd user, the $\varphi_2^{maj}$ is defined as the 2nd ordered channel gain. Consequently, the $\varphi_2^{maj}$ in the $s$th decision set, $\varphi_{2,s}^{maj}$, can be written as in (23). Then, the CDFs $F_{\varphi_{2,0}^{maj}}(x)$ and $F_{\varphi_{2,1}^{maj}}(x)$ can be obtained, respectively as

$$F_{\varphi_{2,0}^{maj}}(x) = F_{x^{(0,1)}|_{l=2}}(x) F_{y^{(0,1)}|_{l=2}}(x)$$
$$= F_{\tilde{\varphi}_{2,0}}(x) F_{\tilde{\varphi}_{2,0}}(x)$$
$$= \left( 6 \left( \frac{1}{2} F_\varphi^2(x) - \frac{1}{3} F_\varphi^3(x) \right) \right)^2$$
$$= 9F_\varphi^4(x) - 12F_\varphi^5(x) + 4F_\varphi^6(x) \qquad (28)$$

$$F_{\varphi_{2,1}^{maj}}(x) = F_{x^{(1,n)}|_{l=2}}(x) F_{y^{(1,n)}|_{l=2}}(x) F_{Z^{(1,n)}}(x)$$
$$= F_{\tilde{\varphi}_{2,1}}(x) F_{\tilde{\varphi}_{2,1}}(x) F_\varphi(x)$$
$$= \left( 2 \left( F_\varphi(x) - \frac{1}{2} F_\varphi^2(x) \right) \right)^2 F_\varphi(x)$$
$$= 4F_\varphi^3(x) - 4F_\varphi^4(x) + F_\varphi^5(x) \qquad (29)$$

Hence, the CDF $F_{\varphi_2^{maj}}(x)$ can be expressed as

$$F_{\varphi_2^{maj}}(x) = P_r(s=0) F_{\varphi_{2,0}^{maj}}(x) + P_r(s=1) F_{\varphi_{2,1}^{maj}}(x)$$
$$= 3F_\varphi^3(x) - \frac{3}{4} F_\varphi^4(x) - \frac{9}{4} F_\varphi^5(x) + F_\varphi^6(x). \qquad (30)$$

For $\varphi_1^{maj}$ and $F_{\varphi_1^{maj}}(x)$: Finally, for the 1st user, its channel gain can be found as the 1st order in the 0th decision set or the minimum channel gain in the 1st decision set. Hence, $\varphi_1^{maj}$ in the $s$th decision set, $\varphi_{1,s}^{maj}$, can be stated as in (24). Then, the CDFs $F_{\varphi_{1,0}^{maj}}(x)$ and $F_{\varphi_{1,1}^{maj}}(x)$ can be obtained, respectively as

$$F_{\varphi_{1,0}^{maj}}(x) = F_{x^{(0,1)}|_{l=1}}(x) F_{y^{(0,1)}|_{l=1}}(x)$$
$$= F_{\tilde{\varphi}_{1,0}}(x) F_{\tilde{\varphi}_{1,0}}$$
$$= \left( 3F_\varphi(x) - 3F_\varphi^2(x) + F_\varphi^3(x) \right)^2$$
$$= 9F_\varphi^2(x) - 18F_\varphi^3(x) + 15F_\varphi^4(x)$$
$$\quad -6F_\varphi^5(x) + F_\varphi^6(x), \qquad (31)$$

$$F_{\varphi_{1,1}^{maj}}(x) = F_{Z^{(1,n)}}(x)$$
$$= 2F_\varphi(x) - F_\varphi^2(x) \qquad (32)$$

Hence, the CDF $F_{\varphi_1^{maj}}(x)$ can be expressed as

$$F_{\varphi_1^{maj}}(x) = P_r(s=0) F_{\varphi_{1,0}^{maj}}(x) + P_r(s=1) F_{\varphi_{1,1}^{maj}}(x)$$
$$= \frac{3}{2} F_\varphi(x) + \frac{3}{2} F_\varphi^2(x) - \frac{9}{2} F_\varphi^3(x)$$
$$\quad + \frac{15}{4} F_\varphi^4(x) - \frac{3}{2} F_\varphi^5(x) + \frac{1}{4} F_\varphi^6(x). \qquad (33)$$

From (27), (30) and (33), $F_{\varphi_l^{maj}}(x)$ can be written in general form as in (11).

ACKNOWLEDGMENT

This work was supported by the Scientific and Technological Research Council of Turkey (TÜBİTAK) with the project number EEEAG-118E274.